\documentclass[twoside,onecolumn]{article}

\usepackage{color}
\usepackage{amsmath}
\usepackage{bm}
\usepackage{graphicx}
\usepackage{float}

\usepackage[english]{babel} 

\usepackage[hmarginratio=1:1,top=32mm,columnsep=20pt]{geometry} 

\usepackage{abstract} 

\usepackage{titlesec} 
\renewcommand\thesection{\Roman{section}} 
\renewcommand\thesubsection{\roman{subsection}} 
\titleformat{\section}[block]{\large\scshape\centering}{\thesection.}{1em}{} 
\titleformat{\subsection}[block]{\large}{\thesubsection.}{1em}{} 

\usepackage{titling} 

\usepackage{hyperref} 

\title{Super-resolution Imaging of the Fluorescent Dipole Assembly with Polarized Structured Illumination Microscopy} 
\author{Karl Zhanghao$^{1,*, \dagger}$, Xingye Chen$^{2,*}$, Wenhui Liu$^2$, Meiqi Li$^1$, Yiqiong Liu$^3$, Yiming Wang$^3$, Sha Luo$^4$, Xiao Wang$^5$, Chunyan Shan$^3$, Hao Xie$^2$, Juntao Gao$^2$, Xiaowei Chen$^4$, Xiangdong Li$^{4,6}$, Yan Zhang$^3$, Qionghai Dai$^{2,\dagger}$, Peng Xi$^{1,\dagger}$\\
$^1$ Department of Biomedical Engineering, College of Engineering, Peking University, Beijing 100871, China\\
$^2$ Department of Automation, Tsinghua University, Beijing 100084, China\\
$^3$ State Key Laboratory of Biomembrane and Membrane Biotechnology, College of Life Sciences, PKU-IDG/McGovern Institute for Brain Research, Peking University, Beijing 100871, China\\
$^4$ State Key Laboratory of Integrated Management of Pest Insects and Rodents, Institute of Zoology, Chinese Academy of Sciences, Beijing 100101, China\\
$^5$ College of Life Sciences, Peking University, Beijing 100871, China\\
$^6$ University of Chinese Academy of Sciences, Beijing 100049, China\\
$^*$ These authors contributed equally to this work.\\
$^\dagger$ Correspondence: P.X. (xipeng@pku.edu.cn), Q.D. (daiqh@tsinghua.edu.cn),  K.Z. (karl.hao.zhang@gmail.com)
}
\date{} 

\begin{document}
\makeatletter
\renewcommand{\maketitle}{\bgroup\setlength{\parindent}{0pt}
\begin{flushleft}
    \huge
    ~\\
  \textbf{\@title}
      ~\\
      ~\\
    \normalsize
  \@author
\end{flushleft}\egroup
}
\makeatother

\maketitle 



\noindent\textbf{Fluorescence polarization microscopy images both the intensity and orientation of fluorescent dipoles, which plays a vital role in studying the molecular structure and dynamics of bio-complex. However, it is difficult to resolve the dipole assemblies on the subcellular structure and their dynamics in living cells with super-resolution. Here we report polarized structured illumination microscopy (pSIM), which decouples the entangled spatial and angular structured illumination through interpreting the dipoles in spatio-angular hyperspace. We demonstrate its application on a series of biological filamentous systems such as cytoskeleton networks and $\lambda$-DNA, and report the dynamics of short actin sliding through myosin-coated surface. Further, pSIM reveals ¡°side-by-side¡± organization of the actin ring structure in the membrane-associated periodic skeleton in hippocampal neurons. It also images the dipole dynamics of green fluorescent proteins labeled to the microtubules in live U2OS cells.  pSIM can be applied directly to a large variety of commercial or home-built SIM systems.}\\

~\\
~\\
\noindent
\textbf{Introduction}

\noindent
The absorption and emission process of fluorophores are both polarization sensitive so that they are modeled as fluorescent dipoles. The dipole model plays an essential role in fluorescence microscopy. On the one hand, the polarization behavior of fluorescent dipoles is highly related to super-resolution imaging. Considering polarized emission, the localization accuracy of single molecule localization microscopy (SMLM) could be greatly improved\cite{1}. With polarization demodulation, the orientation of fluorescent dipoles can also distinguish structures or molecules in sub-diffractive area\cite{2,3,4,5}. On the other hand, the orientation of rigidly connected fluorescent dipoles indicates the molecular orientation of targeted protein or structure, which could be measured by fluorescence polarization microscopy (FPM)\cite{6, 7}. Polarized excitation and detection imaged the 90¡ã rotation of the septin filament organization in living yeast cells during the budding process\cite{8}. Two-photon excitation microscopy with polarization modulation observed G-protein activation on cell membrane\cite{9}. Polarized excitation and two-channel polarized detection investigated the orientation of nucleoporin and supported the ¡®head-to-tail ring¡¯ arrangement of the Y-shaped subcomplex\cite{10}. To sum up, FPM maps the protein structure onto diffraction-limited fluorescence image, revealing their organization and dynamics in a complex cellular environment. 

\noindent
Limited by spatial resolution, these FPM experiments require a simple geometry of the dipole assembly. For example, the orientation of septin filaments was measured only at the beginning or at the end of the budding process when the dipoles on septin filaments have uniform orientation\cite{8}. The organization of septin filaments during the budding period is hard to observe owing to limited spatial resolution compared to their fine structures\cite{11, 12}. Due to the Abbe¡¯s diffraction limit, FPM only obtains an averaged ensemble dipole when multiple dipoles exist within the area of the point spread function (PSF) of the microscope. Limited spatial resolution deteriorates not only underlying subcellular structures but also the accuracy of measured dipole orientation. Therefore, super-resolution FPM is essential for imaging sub-diffraction dipole assemblies with hidden geometry, preferably at high speed to capture the busy dynamics in live cell. 

\noindent
Two categories of super-resolution FPM techniques have been developed recently: polarized SMLM and polarization demodulation. By controlling the labeling sparsity, in vitro experiments enabled polarization analysis of single dipole and resolved the 120¡ã stepping rotation of F1-ATPase\cite{13} and rotational walking of myosin\cite{14,15,16,17,18}, which nevertheless apply to a limited number of bio-complexes and lose in situ information. Combining fluorescence polarization with single molecule localization, polarized SMLM images both the position and orientation of single dipole\cite{19,20,21,22}, which is of vital importance for the studying of protein orientations. However, single molecule localization microscopy requires special sample preparation, and it gains spatial resolution but sacrifices temporal resolution. With minutes to hours for a typical imaging acquisition, it is hard to measure structural dynamics in living cells. Polarization demodulation could obtain super resolution with sparse deconvolution, which achieves fast imaging speed (5 f.p.s) and applies to regular labeling strategies\cite{2,3,4,5}. However, the resolution enhancement of polarization demodulation is dependent on the specimen: super-resolution could be achieved when the specimen has a heterogeneous distribution of dipoles (or sparse polarization signals). Even though Excitation Polarization Angle Narrowing (ExPAN) with stimulated emission depletion (STED) or photoswitchable proteins\cite{23} could further increase the sparsity of polarization signals\cite{2,4} the resolution enhancement would be purely brought by sparse deconvolution if the specimen had a homogeneous distribution of dipoles (no sparsity in polarization signals).

\noindent
Structured Illumination Microscopy (SIM) is suitable for fast live cell imaging with doubled spatial resolution, which reveals abundant subcellular structures and dynamics, including cytoskeleton, mitochondria, endoplasmic reticulum, and intracellular organelle interactions\cite{24,25,26,27}. In a general setup of SIM, linearly polarized lasers interfere to generate structured illumination. Also, polarization modulation is required to obtain high modulation factor, making it a natural fluorescence polarization microscope (Figure 1a). Here we invented polarized SIM (pSIM), which decouples the entangled dipole information in the spatial-angular structured illumination. With careful inspection of the polarization behavior of the SIM system, pSIM maintains the measurement accuracy and sensitivity of the dipole orientation on both home-built and commercial SIM setups, with doubled spatial resolution. Afterward, we primarily applied pSIM on a commercial SIM system, with 2D-SIM, 3D-SIM, or TIRF-SIM imaging modality. pSIM successfully imaged the dipole orientation of cytoskeletal filaments with super resolution in fixed cells, tissue sections, as well as in live cells. 

~\\
~\\
\noindent
\textbf{Results}

\noindent
\textbf{Structured illumination in spatio-angular hyperspace.} To provide a universal framework to model polarization in microscopy, we interpret the specimen in spatio-angular hyperspace\cite{28} or the x-y-$\alpha$ coordinate, by stretching the dipoles on an additional dimension of orientation. The dipoles would be uniformly excited by a circularly polarized light in the angular dimension. In contrast, the dipoles are ¡°structurally illuminated¡± by a linearly polarized light: the dipoles parallel to the polarization have the highest absorption efficiency while the dipoles perpendicular are not excited at all.  Figure 1b illustrates the dipoles in the x-$\alpha$ section of the spatio-angular hyperspace. The quantitative relationship between absorption efficiency and dipole orientation is a cosine-square or sinusoidal function, analogous to spatial structured illumination (Eq. 1). Fourier transform of the sinusoidal function contains three harmonics (0$^{th}$, +1$^{st}$, -1$^{st}$), which could be solved separately by changing the excitation polarization (or changing the phases of angular structured illumination). From the perspective of Fourier space, we explain how polarization modulation enables measurement of dipole orientation by observing additional angular harmonics of dipole orientation information. At least three polarized excitations are required to solve the three harmonics, which is consistent with the perspective of fitting the dipole orientation from its polarization response.

\begin{eqnarray}
\nonumber polarized\quad excitation: \quad & F_\theta(\alpha) = \frac{\eta}{2}[1+cos(2\pi\cdot \frac{1}{\pi} \alpha-2\theta)]\\
structured\quad illumination: \quad & I_{\theta,\phi}(\mathbf{r}) = \frac{I_0}{2}[1+cos(2\pi \mathbf{k_\theta}\cdot \mathbf{r}+\phi)]\\
\nonumber SIM: \quad & D = [S_p(\mathbf{r},\alpha)\cdot I_{\theta,\phi}(\mathbf{r})\cdot F_\theta(\alpha)]\otimes PSF
\end{eqnarray}
\noindent
Here we use $2\pi\cdot\frac{1}{\pi}\cdot\alpha$ to indicate the angular illumination frequency vector ($k_\alpha=\frac{1}{\pi}$) in the same format as structured spatial illumination. $I_{\theta,\phi}(\mathbf{r})$ denotes the structured illumination on spatial dimensions with the stripe direction $\theta$ and the phase $\phi$, $F_\theta(\alpha)$ denotes the absorption efficiency of the fluorescent dipole with orientation $\alpha$, which relates to the excitation polarization of $
\theta$. $D$ denotes the detected image of SIM. $S_p(\mathbf{r},\alpha)$ denotes the specimen in spatio-angular hyperspace. The $\mathbf{k_\theta}$ vector describes the periodicity of the stripes, and its direction is normal to the stripes. $PSF$ denotes the point spread function of the system.

\noindent
SIM generates both spatial structured illumination by interference and angular structured illumination by polarized excitation. Taken the direction of $\mathbf{k_\theta}$ vector as x-axis which is perpendicular to the illumination stripes, we could display the spatio-angular structured illumination in the x-$\alpha$ coordinate (Figure 1c). The spatio-angular pattern of structured illumination contains higher frequency components on all dimensions after Fourier transform (Figure 1e), which would obtain both super resolution and dipole orientation imaging (Details in Supplementary Note 1). We excited the specimen of uniformly distributed 20-nm fluorescent beads with interferometric stripes of two linearly polarized lasers and directly imaged their fluorescent signal in spatial-orientation hyperspace (see Online Method). The experimentally observed illumination pattern and its Fourier transform (Figure 1d, f) are consistent with simulation results.

\noindent
\textbf{Polarized Structured Illumination Microscopy (pSIM).} Applying inverse Fourier transform to all the frequency components, we can achieve super-resolution imaging of both intensity and orientation of fluorescent dipoles. The spatial harmonics (marked in blue) could be solved in the same way as SIM does (Eq. 3). Usually, three directions of interferometric stripes bring six spatial harmonics covering the doubled spatial region on the reciprocal space. Three solved zeroth harmonics from three directions further solves the angular harmonics (marked in yellow, Eq. 4). The spatial harmonics and angular harmonics make up the reciprocal space of pSIM, which extends SIM with additional angular information and extends polarization modulation with doubled spatial resolution. Figure S1 compares the reciprocal space of wide field (WF), SIM, polarization modulation (PM), and pSIM. Applying inverse Fourier transform to the corresponding reciprocal space, the result of WF, SIM, PM, or pSIM is obtained in this paper. The spatio-angular cross harmonics (marked in gray) are unsolvable with the SIM dataset because excitation polarization $\theta$ and illumination vector $\mathbf{k_\theta}$ are dependent on each other. However, the missing harmonics do not influence either spatial image or dipole orientation image according to our simulation (Figure S2).

\noindent
pSIM pays particular attention to the polarization behavior of the SIM system, to guarantee accurate measurement of dipole orientations. We firstly inspect the polarization distortion with a custom-built SIM setup using a spatial light modulator to generate gratings (SLM-SIM), which is widely used to achieve fast SIM imaging\cite{24, 29, 26, 27}(Figure S3a). A polarization beam splitter keeps diffracted laser beams linearly polarized with high extinction ratio, and a vortex half-wave plate rotates the polarization of every illumination pattern. Two identical dichroic mirrors are placed perpendicular to each other on two sides of a 4f system, which cancels their polarization distortion. With all efforts, the extinction ratio of each polarized beam before interference in our system is higher than 10 (Figure S3c). For integrated commercial SIM system, we measure the polarization behavior with the same standard specimen: Phalloidin labeled actin filaments in fixed U2OS cells, whose fluorescence response shows strong fluctuation under different  polarization excitation (Supplementary Movie 1).  The polarization of the laser beams could be inferred indirectly from the modulation factor of interferometric stripes or the polarization factor of the actin filaments. We examined a GE OMX SR system (OMX-SIM) and a Nikon SIM system (N-SIM), which have comparable polarization performance compared to home-built SLM-SIM on the test slides (Figure S4).

\noindent
In the conventional polarization modulation system, only the polarization of the excitation beam is rotated. Whereas, both the polarization and the interferometric stripes are rotated during SIM imaging, which would introduce intensity non-uniformity among three directions. To calibrate the intensity non-uniformity, we use sparsely distributed 100-nm fluorescent beads, whose fluorescence is constant under different polarizations so that their fluorescent signal reflects only the intensity non-uniformity of the system. Usually, the inner area of the field of view (FOV) has small non-uniformity while the outer area may have a non-uniformity as large as 50\% (Figure S4, 5). After calibration of the intensity non-uniformity, pSIM could achieve accurate measurement of the dipole orientation in the whole FOV (Online Method, Figure S5).

\noindent
With polarization distortion compensated and intensity non-uniformity calibrated, pSIM can measure the dipole orientation as accurately as conventional polarization modulation methods. We imaged SYTOX Orange labeled lambda DNA filaments, into which the fluorescent dipoles perpendicularly insert\cite{20}. The dipole orientation is marked with pseudocolor, with a color wheel indicating the relationship between the color and the orientation (Figure 2c-e). The orientation deviation from the normal of the filament measured is 3.85$^\circ$ in pSIM results (Figure 2f). Compared to polarization modulation, pSIM doubles the highest observed spatial frequency, which can measure dipole orientations in together with super-resolution observation of the underlying structures. We simulated two tangled lines with dipoles parallel to each line and compare the imaging results between polarization modulation and pSIM, the latter of which indicates the underlying structure and the dipole orientations (Figure 2g, h). As for the intersecting area or wobbling dipoles illustrated elsewhere\cite{22}, pSIM results indicate the ensemble orientation of the dipoles. In this paper, we measure the average orientation of the dipole assembly on the cytoskeletal filament, which reflect the structural information of the filaments.

\noindent
\textbf{Imaging results of actin filaments in fixed cells.} We imaged phalloidin-AF488 labeled actin filaments in fixed BAPE cells with 2D SIM on OMX-SIM. Figure 3a, b displays the corresponding WF, SIM, PM, and pSIM results. pSIM provides the same spatial resolution as SIM, which is doubled compared to WF or PM results. PM and pSIM measure the ensemble orientation of the AF488 dipoles, which is mostly parallel to the direction of filaments. Hence, the color wheel also indicates the relationship between the color and the filament direction. The orientation deviation from the tangent direction of the filament is 8$^\circ$ with pSIM results, indicating the accuracy of pSIM on measuring dipole orientation. For thick specimens, we combined pSIM with 3D-SIM, which additionally achieves axial super-resolution and optical sectioning\cite{30, 31}. Figure 3g, h shows the Maximum Intensity Projection (M.I.P.) image of actin filaments of a mouse kidney tissue section (also Supplementary Movie 2). The actin protein forms much thicker actin fiber bundle in the brush rim of the kidney nephric tubule, with parallel polarization to the actin filaments. 3D-pSIM yields much sharper details of the actin fiber bundles, while the PM result is blurry with the out-of-focus background. With TIRF-SIM imaging modality, we imaged phalloidin-AF568 labeled actin filaments in U2OS cells embedded in PBS buffer and compared the results of TIRF-PM and TIRF-pSIM  (Figure S6).

\noindent
Recent works using cryo-electron microscopy\cite{32} and AFM\cite{33} revealed that eukaryotic cells contain both long ($>$5 $\mu m$) and short ($<$0.5 $\mu m$) actin filament. The short actin filament is hard to resolve for conventional microscopy, let alone distinguishing its orientation from intensity image. We examined the capacity of pSIM in determining the orientation of short actin filaments by tracking the in-vitro actin filaments gliding on the myosins. For long actin filaments, the gliding direction is consistent with the tangent direction of the actin filaments, as well as the dipole orientation measured by pSIM. For the short actin filament whose direction is indistinguishable, the gliding direction is consistent with the measured dipole orientation (Figure 4a-c, Supplementary Movie 3-4).

\noindent
In the membrane-associated periodic skeleton (MPS) recently discovered in neurons\cite{34,35,36}, short Adducin-capped actin filaments form quasi-1D periodical ¡°actin ring¡± structure, which is an essential building block of MPS. Either STORM or STED is capable of revealing the actin ring structure, which is transverse to the MPS direction. However, even the cutting-edge super-resolution techniques fail to observe the organization of the actin filaments, which are tens of nanometer long segments and densely packed, hidden behind the resolving capability. The ¡°end-to-end¡± organization of actin filaments was assumed and supported by indirect evidence, which shows the diameter increase of axons after depletion of Adducin\cite{37}. pSIM can distinguish the continuous long actin filaments in the dendrites and the discrete ¡°actin ring¡± in the axons. The ¡°actin ring¡± structure in the 2D-pSIM section displays 184-nm periodicity after Fourier transform (Figure 4i), which is consistent with previous STORM or STED results. The direction of short actin filaments inferred from additional polarization information is indeed parallel to the MPS structure, which rejects the ¡°end-to-end¡± model. Instead, the ¡°side-by-side¡± organization of the ring structure is in agreement with pSIM result (Figure 4k). PM result may also reveal the parallel polarization of actin filaments; nevertheless it will fail to resolve the discrete ¡°ring structure¡±. The unique organization of MPS may play a vital role in the neuronal structure and consequently its plasticity. pSIM imaging results add significant details to the existing MPS model, since different organization of the actin filaments may result in the differentiated functionality of MPS. Such a parallel package of actin filaments makes these structures more flexible to relocation and movement along the axon. 

\noindent
\textbf{Imaging results of GFP labeled microtubule in live U2OS cells.} 3D-pSIM images the microtubule in live U2OS cells expressing $\alpha$-tubulin-GFP (Figure 5a, b), and obtains time-lapse imaging of the microtubules (Figure 5c, Supplementary Movie 5). For dynamic imaging, the imaging speed should be faster than the movement of the specimen; otherwise, the motion would cause not only the blurry image but also incorrect orientation measurement. Here we obtain volumetric imaging of the microtubules at a speed of 0.67 reconstructed f.p.s. with 3-ms exposure time. The ensemble dipole orientation on the microtubule is mostly perpendicular to the filament in all images. The ensemble dipole on the microtubule consists of the 2D in-plane projection of all the 3D GFP dipoles linked to the $\alpha$-tubulin subunit. Due to the central symmetry of the microtubule, the ensemble dipole can only be perpendicular or parallel to the filament. When the included angle between the GFP dipole and the filament is close to 90$^\circ$ (or 0$^\circ$), the ensemble dipole would be perpendicular (or parallel) to the filament (Figure 5d-f). The included angle in between or the wobbling behavior of the single GFP will lead to an ensemble dipole with smaller polarization factor (smaller polarization response).

~\\
~\\
\noindent
\textbf{Discussion}

\noindent
Although powerful in resolving the dipole orientation, the conventional FPM techniques were previously obstacled by poor spatial resolution due to diffraciton limit. Here we introduce pSIM, which can map the dipole onto spatial-angular hyperspace to obtain polarization information from SIM instrument instantly. Its broad applicability has been demonstrated with different types of specimens, such as ¦ËDNA, actin filament in BAPE cells and mouse kidney tissue, interaction between actin and myosin, as well as microtubules stained with GFP in live U2OS cell. Moreover, the ¡°side-by-side¡± model of the MPS in axon has been successfully interpreted. Along these experiments, we have demonstrated that pSIM is compatible with a variety of SIM modalities such as 2D-SIM, TIRF-SIM, and 3D-SIM. TIRF-SIM uses two first-order diffracted beams as 2D-SIM does, with the incident angle larger than the critical angle. Since the two beams contain only s-polarization, the polarization would change under TIRF condition. High-NA TIRF-SIM, which gains a resolution of 86 nm, should also be compatible with pSIM since the s-polarization maintains in the system\cite{24}. Thanks to the removal of out-of-focus background and high signal-to-noise ratio, TIRF-SIM has achieved ultrafast imaging speed with SLM switching the diffractive patterns\cite{27}. Grazing incidence SIM (GI-SIM) extends the depth-of-focus of ~100 nm in TIRF to ~1 $\mu$m with minimal out-of-focus background\cite{26}, which should also be compatible with pSIM. 3D-SIM includes the zeroth-order diffracted beam to generate a 3D interferometric pattern, which adds axial dimension to the reciprocal space of 3D-pSIM. The polarization behavior of 3D-SIM is the same with 2D-SIM. However, only 2D in-plane dipole orientation is measured in 3D-pSIM since only s-polarized excitation exists. Both OMX-SIM and N-SIM have the capability of 2D-SIM, TIRF-SIM, and 3D-SIM, while SLM-SIM can perform 2D-SIM and TIRF-SIM. One should be informed of its polarization behavior before applying pSIM to existing SIM systems. Indeed, pSIM is incompatible with those using the incoherent light source\cite{38}, and it fails with instant SIM\cite{39} or other setups without polarization modulation\cite{40}.

\noindent
The performance of pSIM is dependent on the SIM system. Typically, the lateral resolution is \textasciitilde 100 nm, and the axial resolution is \textasciitilde 300 nm. The fastest imaging speed of SLM-SIM, OMX-SIM, and N-SIM are \textasciitilde 100, 15, and 0.7 fps respectively for 2D imaging. Polarized SMLM has much higher spatial resolution and measures the orientation of single dipoles, which may be capable of resolving the underlying structure within the dipole assembly. However, special sample preparation and long acquisition time make it impossible to image the dynamics in live cells. pSIM could be a complementary technique, which images high-order organization of these dipole assemblies and captures their dynamics. pSIM has no restriction on fluorescent labeling so that it applies to a large variety of specimen. The general compatibility of pSIM with 3D-SIM or TIRF-SIM makes it suitable for imaging either thick specimen or specimen near the coverglass. Moreover, biologists can readily perform pSIM on existing commercial systems, which will instantly advance studying the structure and dynamics of bio-complexes.

~\\
~\\
~\\
\noindent \textbf{Online Methods} \\

\noindent \textbf{Reconstruction algorithm of pSIM.} pSIM imaging process is described by Eq.1, whose Fourier transform would be:
\begin{eqnarray}
\nonumber  &  \tilde{D}_{\theta,\phi}(\mathbf{k_r},k_\alpha)=[\tilde{S}_p(\mathbf{k_r},k_\alpha)\otimes\tilde{I}_{\theta,\phi}(\mathbf{k_r})\otimes\tilde{F}_\theta(k_\alpha)]\cdot OTF(\mathbf{k_r},k_\alpha)\\
& \tilde{I}_{\theta,\phi}(\mathbf{k_r}) = \frac{\pi I_0}{4}[\delta(\mathbf{k_r})+\frac{1}{2}e^{i\phi}\delta(\mathbf{k_r}-\mathbf{k_\theta})+\frac{1}{2}e^{-i\phi}\delta(\mathbf{k_r}+\mathbf{k_\theta})]\\
\nonumber & \tilde{F}_\theta(k_\alpha)=\frac{\pi\eta}{4}[\delta(k_\alpha)+\frac{1}{2}e^{-2i\theta}\delta(k_\alpha-\frac{1}{\pi})+\frac{1}{2}e^{2i\theta}\delta(k_\alpha+\frac{1}{\pi})]
\end{eqnarray}

\noindent
Both the spatial structured illumination, $I_{\theta,\phi}(\mathbf{r})$ and orientational structured illumination $F_\theta(\alpha)$ bring larger observable region on spatial dimensions and orientational dimension. The reconstruction of PSIM takes two steps: ¡°SIM step¡± and ¡°PM step¡±. For ¡°SIM step¡±, three images belong to the same pattern are used to solve the three frequency components as conventional SIM does. However, every frequency
component is convoluted with a polarization term $\tilde{F}_\theta(k_\alpha)$, as shown in Eq. 3

\begin{eqnarray}
\begin{bmatrix}
\tilde{D}_{\theta_i,\phi_1}(\bm{k_r},k_\alpha)\vspace{1ex}\\
\tilde{D}_{\theta_i,\phi_2}(\bm{k_r},k_\alpha)\vspace{1ex}\\
\tilde{D}_{\theta_i,\phi_3}(\bm{k_r},k_\alpha)\vspace{1ex}\\
\end{bmatrix}&=&\textcolor[rgb]{0,0,1}{M_{SIM}}\cdot\begin{bmatrix}
[\textcolor[rgb]{0,0,1}{\tilde{S}_p(\bm{k_r},k_\alpha)}\otimes \textcolor[rgb]{1,0,1}{\tilde{F}_{\theta_i}(k_\alpha)}]\cdot OTF(\bm{k_r},k_\alpha)\vspace{1ex}\\
[\textcolor[rgb]{0,0,1}{\tilde{S}_p(\bm{k_r-k_{\theta_i}},k_\alpha)}\otimes \textcolor[rgb]{1,0,1}{\tilde{F}_{\theta_i}(k_\alpha)}]\cdot OTF(\bm{k_r},k_\alpha)\vspace{1ex}\\
[\textcolor[rgb]{0,0,1}{\tilde{S}_p(\bm{k_r+k_{\theta_i}},k_\alpha)}\otimes \textcolor[rgb]{1,0,1}{\tilde{F}_{\theta_i}(k_\alpha)}]\cdot OTF(\bm{k_r},k_\alpha)\vspace{1ex}\\
\end{bmatrix};\\
\nonumber \textcolor[rgb]{0,0,1}{M_{SIM}}&=& \frac{\pi I_0}{4}\begin{bmatrix}
1 &\frac{1}{2}e^{i\phi_1}&\frac{1}{2}e^{+i\phi_1}\vspace{1ex}\\
1 &\frac{1}{2}e^{i\phi_2}&\frac{1}{2}e^{+i\phi_2}\vspace{1ex}\\
1 &\frac{1}{2}e^{i\phi_3}&\frac{1}{2}e^{+i\phi_3}
\end{bmatrix} 
\end{eqnarray}

\noindent
Then ¡°PM step¡± follows. From the 3 original spatial components $\tilde{S}_p(\mathbf{k_r},k_\alpha)\otimes\tilde{F}_{\theta_i}(k_\alpha), (i=1,2,3)$, $\tilde{S}_p(\mathbf{k_r},k_\alpha)$, $\tilde{S}_p(\mathbf{k_r},k_\alpha-\frac{1}{\pi})$, $\tilde{S}_p(\mathbf{k_r},k_\alpha+\frac{1}{\pi})$ could be further solved with PM equation (Eq. 4). SIM uses 3 directions of illumination patterns, which cover the doubled region in reciprocal space. The 3 directions of polarization are just sufficient to extract the dipole orientations, while PM systems usually use more excitation polarizations to obtain robust results. Before the ¡°PM step¡±, the images are compensated with the calibration data of fluorescent beads.

\begin{eqnarray}
\begin{bmatrix}
\textcolor[rgb]{0,0,1}{\tilde{S}_p(\bm{k_r},k_\alpha)}\otimes \textcolor[rgb]{1,0,1}{\tilde{F}_{\theta_1}(k_\alpha)}\vspace{1ex}\\
\textcolor[rgb]{0,0,1}{\tilde{S}_p(\bm{k_r},k_\alpha)}\otimes \textcolor[rgb]{1,0,1}{\tilde{F}_{\theta_2}(k_\alpha)}\vspace{1ex}\\
\textcolor[rgb]{0,0,1}{\tilde{S}_p(\bm{k_r},k_\alpha)}\otimes \textcolor[rgb]{1,0,1}{\tilde{F}_{\theta_3}(k_\alpha)}\vspace{1ex}\\
\end{bmatrix}&=& \textcolor[rgb]{1,0,1}{M_{PM}}\cdot\begin{bmatrix}
\textcolor[rgb]{1,0,0}{\tilde{S}_p(\bm{k_r},k_\alpha)}\cdot OTF(\bm{k_r},k_\alpha)\vspace{1ex}\\
\textcolor[rgb]{1,0,0}{\tilde{S}_p(\bm{k_r},k_\alpha-\frac{1}{\pi})}\cdot OTF(\bm{k_r},k_\alpha)\vspace{1ex}\\
\textcolor[rgb]{1,0,0}{\tilde{S}_p(\bm{k_r},k_\alpha+\frac{1}{\pi})}\cdot OTF(\bm{k_r},k_\alpha)\vspace{1ex}\\
\end{bmatrix};\\
\nonumber \textcolor[rgb]{1,0,1}{M_{PM}}&=&\frac{\pi\eta}{4}\begin{bmatrix}
1 & \frac{1}{2}e^{-i2\theta_1}&\frac{1}{2}e^{+i2\theta_1}\vspace{1ex}\\
1 & \frac{1}{2}e^{-i2\theta_2}&\frac{1}{2}e^{+i2\theta_2}\vspace{1ex}\\
1 & \frac{1}{2}e^{-i2\theta_3}&\frac{1}{2}e^{+i2\theta_3}\\
\end{bmatrix} 
\end{eqnarray}

\noindent
These 3 components solved in ¡°PM step¡±, together with other 6 components ($\textcolor[rgb]{1,0,0}{\tilde{S}_p(\bm{k_r\pm k_{\theta_i}},k_\alpha)\otimes \tilde{F}_{\theta_i}(k_\alpha)},i=1,2,3$) solved in ¡°SIM step¡± make up the observable region of reciprocal
space of PSIM (Figure 1g). The six high-order spatial components ($\tilde{S}_p(\mathbf{k_r}\pm \mathbf{k_{\theta_i}},k_\alpha)\otimes\tilde{F}_{\theta_i}(k_\alpha),i=1,2,3$) could not be further solved to get separated polarization components $\tilde{S}_p(\mathbf{k_r}\pm\mathbf{k_{\theta_i}},k_\alpha)$ and $\tilde{S}_p(\mathbf{k_r}\pm\mathbf{k_{\theta_i}},k_\alpha\pm\frac{1}{\pi}),i=1,2,3$ (cross first harmonics of frequency components). Assembling the nine shifted components in reciprocal space and applying inverse Fourier transform on spatial dimensions. For PM and pSIM, the polarization response on each pixel is fitted to the equation $I=Acos(2(\theta-\alpha))+B$, with $\theta$ denoting polarization and $I$ denoting the image intensity. We use $A+B$ as the intensity signal and $\alpha$ as the dipole orientation. We define $\frac{2A}{A+B}$ as the polarization factor, which equals the definition elsewhere\cite{21}. Detailed derivation of pSIM is in Supplementary Note 2.

\noindent
Our SIM reconstruction algorithm is based on previous work of fairSIM (https://www.fairsim.org/)\cite{41}, which is an ImageJ plugin written in Java. However, our data is analyzed by home-written Matlab program for easier debugging. To facilitate the science community, we have released our source code on Github (https://github.com/chenxy2012/PSIM).

\noindent
\textbf{Calibration of illumination fluctuation.} The slide of 100 nm fluorescent beads is prepared at a proper density so that the beads can be localized separately. The beads are imaged by 2D SIM sequence at largest field-of-view of the system. For each pattern, three images of three different phases calculate the wide field image (the 0$^{th}$ spatial harmonic). If the phase difference is designed to be $2\pi/3$, wide field image could be easily obtained by averaging the three images. In each wide field image, the beads are localized by QuickPALM (http://code.google.com/p/quickpalm)\cite{42} with their position and intensity exported. For those beads appears on all three images at the same position, they are used to compensate illumination non-uniformity among different patterns. We either use a quantic polynomial function to fit the non-uniformity or move the beads at a step size of 500 to calibrate the whole field-of-view. Compensation of intensity non-uniformity is performed before the ¡°PM step¡± during PSIM reconstruction based on Eq. 5. Detailed information is in Supplementary Note 3.

\begin{eqnarray}
D &=& (S\cdot Calib)\otimes PSF\\
\nonumber D_{calib} &=& F^{-1}\{ F\{ D\}/OTF \}/Calib
\end{eqnarray}

\noindent
\textbf{Experimental measurement of structured illumination.} A single-layer uniform 20 nm fluorescent beads sample is prepared to measure the spatio-angular structured illumination, which is excited by structured illumination in our home-built SLM-SIM system. A polarizer in front of the detector rotates from 0 to 180 degrees, and the images are captured every 20 degrees. All images are rotated for the same angle to make the stripes vertical. Afterward, columns of the image are averaged to form a row as a spatial dimension. The polarization signals are placed in a column forming the angular dimension (Figure 1d). Finally, the frequency domain of the structured illumination in spatio-angular hyperspace is acquired by 2D Fourier Transform (Figure 1f).

\noindent
\textbf{Simulations to verify PSIM.} The spatio-angular cross harmonic frequency components are unresolvable by pSIM so that pSIM uses the obtained nine frequency pedals to generate dipole orientation image. If all the components are solved separately, 21 frequency pedals are obtained to fill up the full doubled region (Figure S2). To study the influence of missing cross harmonic frequencies, we simulated radial lines and circles with parallel dipole orientation to their direction. The simulated specimen in x-y-$\alpha$ coordinate is discretized with a spatial grid of 20 nm and an angular grid of 12.5 degree. Then we apply Fourier transform to the simulated data and obtain corresponding frequency pedals. Frequency components beyond the observable area of pSIM (Figure S1a) or full double region are neglected (Figure S2b). Afterward, inverse Fourier transform is applied to obtain super-resolution dipole imaging. We find that the missing frequency components do not influence either intensity image or dipole orientations. They only influence the polarization ratio of the specimen\cite{3}, which describes the variation of dipole orientations or wobbling of dipoles on each pixel. Unexpected high-frequency fluctuations would appear on the super-resolution image of polarization ratio (Figure S2g).

\noindent
\textbf{SIM setup and imaging.} Figure S3 shows the schematic setup of the home-built SLM-SIM. A laser beam  (CNI, MGL-FN-561-200 mW) was expanded by an achromatic beam expander (Thorlabs, GBE10-B). A half-wave plate (Union, WPA2420-450-650) adjusts the polarization before the laser passes through a polarization beam splitter (Thorlabs, CCM1-PBS251). The ferroelectric liquid crystal spatial light modulator (SLM; Forth Dimension Displays, SXGA-3DM-DEV) controls the angle and the phase of the diffraction pattern. Other orders except the $\pm$1 diffraction orders were blocked by a spatial filter (mask). We use a vortex half-wave plate (Thorlabs, WPV10L-532) to modulate the polarization of the two $\pm$1 order beams to be parallel to the interference stripes. A dichroic mirror DM1 (Chroma, ZT561rdc), placed in perpendicular orientation to the DM2, was introduced to compensate the polarization ellipticity by switching the incident position of s-beam and p-beam. A lens pair were used to relay ¡À1 order light spots to the back focal plane of the objective (Nikon, CFI Apochromat TIRF 100$\times$ oil, NA 1.49). The fluorescent signals pass the emission filter (Semrock, FF01-640/20-25), and reach a sCMOS camera (Tuscen, Dhyana 400BSI).
pSIM on the commercial OMX-SIM system (DeltaVision OMX SR, GE, US) uses 60$\times$ 1.4 NA oil immersion objective (Olympus, Japan) and AF488 or AF561 filter sets. Standard 2D SIM or 3D SIM sequence was performed with 80 nm pixel size and 125 nm axial step. pSIM on commercial N-SIM system (Nikon, Japan) uses 100$\times$ 1.49 NA oil immersion objective (Apo TIRF, Nikon, Japan). The 2D SIM was performed with 65 nm pixel size.

\noindent
\textbf{Sample preparation.} The specimen of Phalloidin-AF488 labeled F-actin in BAPE cell and Phalloidin-AF568 labeled actin in mouse kidney section are commercially available (FluoCells Prepared Slide \# 1 and FluoCells Prepared Slide \# 3, ThermoFisher). U2OS cells were cultured in DMEM medium and 10\%  (v/v) FBS at 37 $^\circ$C, 5\% CO2, on 0.17-mm coverslips. Primary culture neurons were cultured from the hippocampus of newborn C57 mice. The use of mice was in accordance with the regulations of the Peking University Animal Care and Use Committee. Fetal hippocampal samples were dissociated from the brain and digested by 0.25\% trypsin (Invitrogen). Digestion was stopped by adding DMEM-F12 medium (Gibco) with 10\% FBS (Gibco), after which the tissue was dispersed by a pipette. After 2 minutes of precipitation, the supernatant was collected and centrifuged at 500 $\times$ g for 2 minutes. Afterward, the cells were resuspended in DMEM-F12 medium with 10\% FBS and plated on a coverslip coated with poly-D-lysine (Sigma) in 5\% circulating CO2. Neurobasal medium (Gibco) containing Pen-Strep (Invitrogen), B27 (Gibco), and GlutaMAX (Thermo Fisher) was added to the medium after 4 hours. Half of the medium was replaced with fresh medium every 3 days. 
\noindent
For immunostaining, the cells were washed in PBS and fixed in 4\% PFA (Sigma) at room temperature. Then cells were permeabilized in 0.1\% Triton at 4 $^\circ$C, after which they were blocked in 5\% donkey serum at room temperature. Phalloidin-AF568 (A12380, Invitrogen) was added to label actin filament for 1 hour. The cells were washed and mounted on regular slides with Prolong Diamond (P36970, Invitrogen) unless otherwise indicated. For GFP labeling, the plasmid of tubulin-GFP was transfected into U2OS cells under the standard protocol of Lipofectamine 3000 (L3000, Invitrogen).

\noindent
For $\lambda$-DNA, microscope coverslips were first covered with a thin layer of PMMA (Poly-methyl methacrylate) to stretch DNA onto the coverslips. SYTOX Orange Nucleic Acid Stain (5 mM Solution in DMSO, Invitrogen) were diluted to 1000 times. Then 0.3 $\mu l$ stock $\lambda$-DNA solution (300 $\mu$g/$\mu$l, Invitrogen) was dissolved in 968 $\mu$l PBS, and 32 $\mu$l diluted sytox were mixed. After that, 5 $\mu$l of the mixed solution is divided into nine drops at the coverslips. Coverslips were allowed to air-dry for almost 1 hour and were sealed with Fixogum rubber cement.

\noindent
The in vitro actin gliding assays were performed using full-length smooth muscle myosin (SmM-FL) and rabbit striated muscle actin \cite{43}. Approximately 20 $\mu$L of 0.4 mg/mL SmM-FL in Rigor solution [25 mM imidazole hydrochloride (pH 7.5), 25 mM KCl, 5 mM MgCl2, and 1 mM EGTA] was introduced into a nitrocellulose-coated flow chamber and incubated for 10 min on ice. After being blocked with 20 $\mu$L of 1 mg/mL BSA in Rigor solution on ice for 5 min, the flow chamber was incubated with 20 $\mu$L of phosphorylation buffer (5.5 $\mu$M CaM, 1.3 $\mu$M myosin light chain kinase, 0.2 mM CaCl2, 5 mM ATP, 1 mM DTT, and 5 nM unlabeled F-actin in Rigor solution) at 25 $^\circ$C for 10 min. The flow chamber was washed with 20 $\mu$L of 1 mg/mL BSA in Rigor solution to remove the unbound proteins and then incubated with 20 $\mu$L of 5 nM Alexa Fluor 488-phalloidin-labeled F-actin in actin gliding buffer I (2.5 mg/mL glucose, 2 units/mL catalase, 40 units/mL glucose oxidase in Rigor solution) on ice for 5 min. The unbound F-actin was washed away with 20 ¦ÌL of motility buffer I. Before observation, the flow chamber was perfused with 20 ¦ÌL of motility buffer II (0.5\% methylcellulose and 5 mM ATP in motility buffer I). 

\noindent
\textbf{Statistics and reproducibility.}  All the figures show the representative data from $\ge$ 3 representative experiments. The fitting curves in Figure 2f and Figure 3f are generated using the Gaussian fitting function in Matlab. The standard deviation of the dipole orientation angles in Figure 2f and Figure 3f were analyzed with Matlab.

~\\
~\\
\noindent 
\textbf{Author Contributions} \\
\noindent 
KZ conceived the project. PX and QD supervised the research. XC and KZ programed reconstruction algorithm and analyzed the data. WL, ML, and KZ built the SLM-SIM system. YL, YW, and KZ performed neuron experiments. ML, XC, and SL performed in vitro actin and $\lambda$-DNA experiments. KZ, PX, XC, and ML wrote the manuscript with inputs from all authors.

~\\
~\\
\noindent 
\textbf{Acknowledgments}\\
This work was supported by the National Key Research and Development Program of China (2017YFC0110202), the National Natural Science Foundation of China (61475010, 61729501, 61327902), the Distinguished Young Scholars of Beijing supported by Beijing Natural Science Foundation, Clinical Medicine Plus X-Young Scholars Project of PKU, and Innovative Instrumentation Fund of PKU. KZ acknowledges the support from China Postdoctoral Science Foundation. We thank the Core Facilities of Life Sciences, Peking University for assistance with SIM Imaging.

\newpage
\begin{figure}[H]
\includegraphics[width=5in]{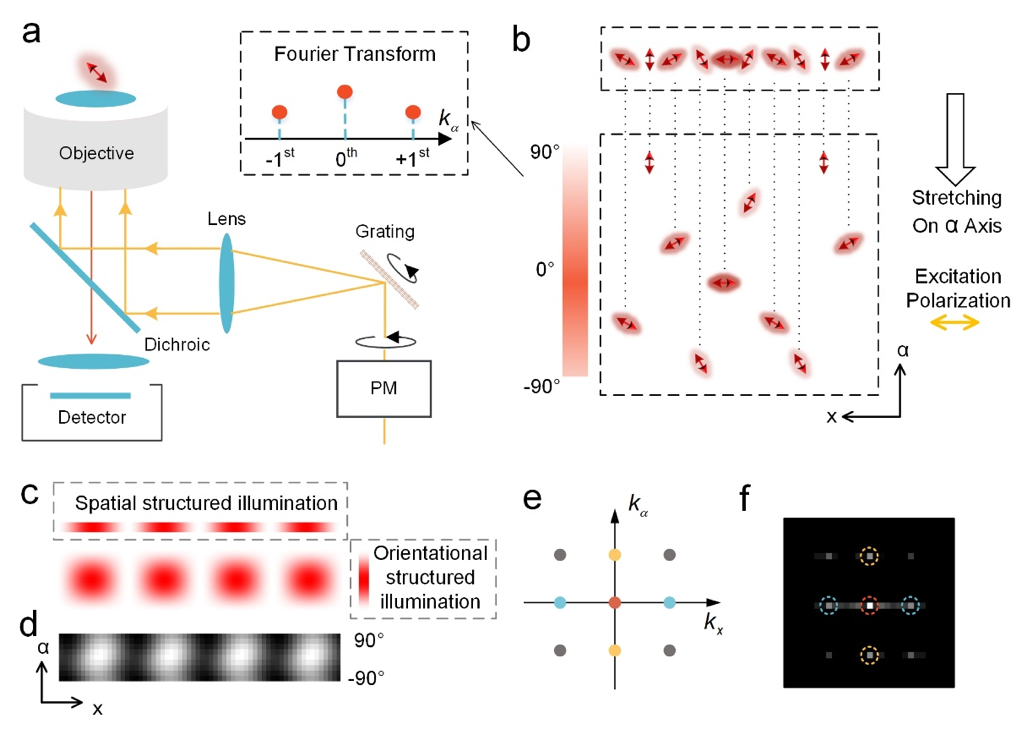}
\caption{\textbf{Principle of polarized Structured Illumination Microscopy (pSIM).} (a) A schematic setup of a typical SIM system. The excitation polarization rotates with the grating to keep laser beams s-polarized, obtaining high-contrast interferometric stripes. (PM: polarization modulation). (b) By stretching the fluorescent dipoles on additional orientation dimension, we interpret them in spatio-angular hyperspace. A linearly polarized light (horizontal polarization) would excite the dipoles with different orientations in a structured manner on angular dimension. The quantitative relationship is a cosine-square function, whose Fourier transform contains three harmonics. From this perspective, polarized excitation is intrinsically structured illumination on angular dimension. (c) Under the illumination of interferometric stripes generated by s-polarized laser beams, the specimen is structurally illuminated on both spatial and angular dimension. Eq. 1 quantitatively describe the spatial structured illumination, the angular structured illumination, and the 2D illumination pattern. (d) We excited uniformly distributed 20-nm fluorescent beads with polarized structured illumination and used a rotary polarizer before the sensor to directly image the illumination pattern in the x-$\alpha$ coordinate, which is consistent with the simulation result. (e) Fourier transform of the 2D illumination pattern in the x-$\alpha$ coordinate brings spatial harmonics (blue), angular harmonics (yellow), and cross harmonics (gray).  (f) is the Fourier transform of the experimental 2D structured illumination in (d), and the corresponding harmonics are marked by the colored circles. }
\end{figure}
\newpage
\begin{figure}[H]
\includegraphics[width=5in]{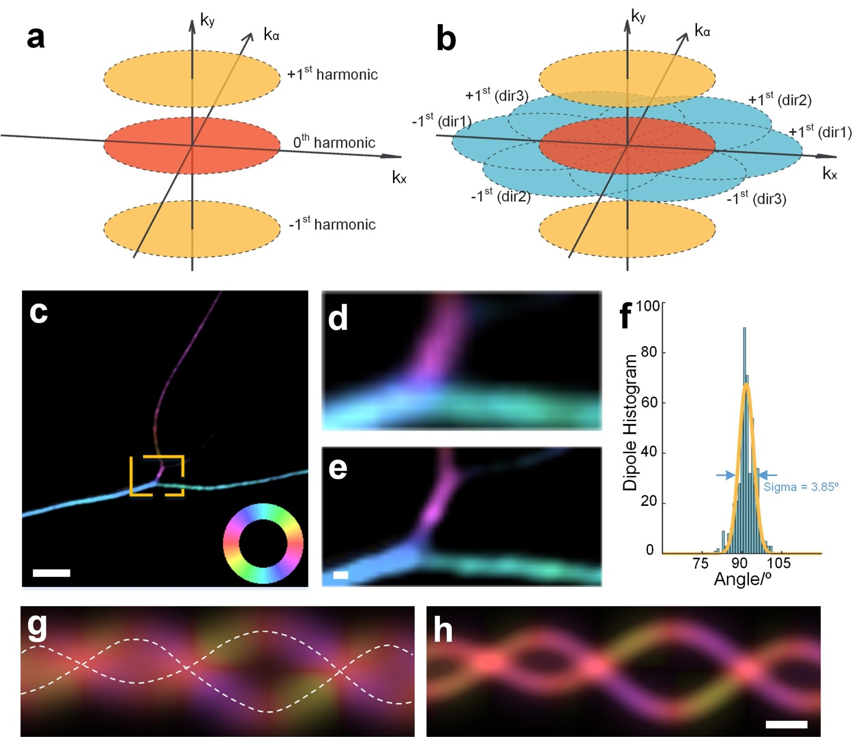}
\caption{\textbf{Comparison between polarization modulation and pSIM.} (a) displays the reciprocal space of polarization modulation microscopy, including three harmonics. The zeroth harmonic determines the intensity image, while the phase of the first harmonics determine the dipole orientation. (b) Three directions of structured illumination bring six spatial first harmonics (blue), which make up the reciprocal space of pSIM in together with the zeroth harmonic (red) and two angular first harmonics (yellow). (c) pSIM imaging results of SYTOX orange labeled DNA filaments, with the dipole orientation marked in pseudocolor. The color wheel indicated the relationship between the dipole orientation and the pseudocolor. The dipole orientation of SYTOX orange is inserted perpendicular into the DNA filament. The zoomed-in results in (d, e) compares the results of polarization modulation and pSIM. (f) is the histogram of the dipole orientation in (c), where the angle represents the difference between the dipole orientation and the direction of the DNA filament. (g, h) Two filaments are simulated, with the dipole orientation tangent to the filament. Compared with polarization modulation (g), pSIM result (h) shows the underlying structure and measures the dipole orientation simultaneously. Scale bar: (c) 2 $\mu$m, (d-h) 200 nm.}
\end{figure}
\newpage
\begin{figure}[H]
\includegraphics[width=5in]{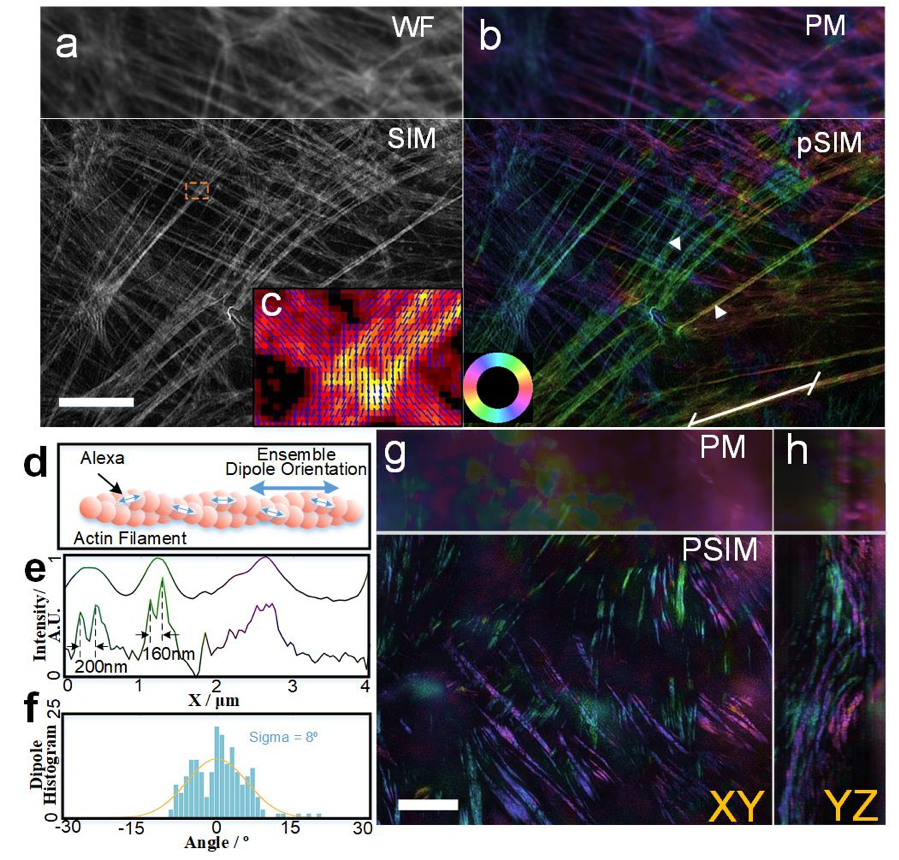}
\caption{\textbf{PSIM imaging results.} (a, b) 2D-pSIM images Phalloidin-AF488 labeled actin in BPAE cells. (a) is the intensity image, in which the upper part is the wide field (WF) result and the lower part is the SIM result. (b) uses pseudocolor to mark the orientation of dipoles. The lower part is the pSIM result, which achieves super-resolution compared to the polarization modulation (PM) result and obtains dipole orientations compared to SIM. A color wheel on the bottom-left indicates the relationship between pseudocolor and dipole orientation. (c) Magnified view of the boxed region in (a) with the dipole orientations shown by the blue arrows. (d) The schematic of Phalloidin-AF488 labeled actin filament, on which the ensemble dipole orientation is parallel to the filament. (e) The intensity profile of PM and pSIM between the two arrows in (b), with pseudocolor indicating their corresponding dipole orientation. PSIM reveals the dipole orientation on individual actin filaments. (f) The dipole histogram of the white line in b. The angle represents the difference between dipole orientation and filament direction. (g, h) 3D-pSIM images the Phalloidin-AF568 labeled actin in mouse kidney section. The Max Intensity Project (M.I.P.) images on the XY plane and the YZ plane are compared. The pseudo-colors used in b, g, h share the same color-wheel. Scale bar: 5 $\mu$m.}
\end{figure}
\newpage
\begin{figure}[H]
\includegraphics[width=5in]{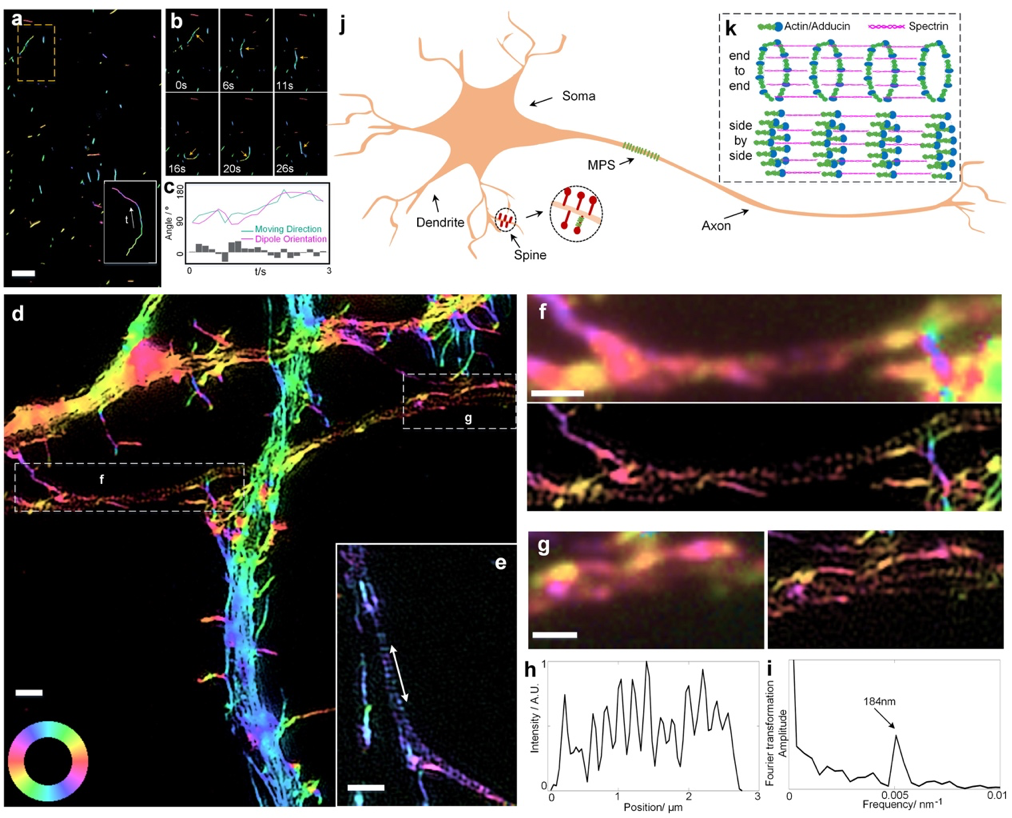}
\caption{\textbf{ Imaging the orientation of short actin filaments.}(a) Dynamic imaging of the myosin-driven movement of Phalloidin-AF568 labeled actin. The white box contains the trajectory of a short actin filament. (b) Magnified view of the yellow boxed region in (b). The dipole orientation of the actin filaments changes during their movement. (c) Time-lapse orientation and position of the fragmented actin in (h). (d) 2D-pSIM imaging of Phalloidin-AF568 labeled actin filaments in hippocampus neurons, which clearly distinguishes the continuous long actin filaments in the dendrite and the section of the discrete ¡°actin ring¡± structure in the axon. (e) Another illustration of ¡°actin ring¡± structure in the axon. (f, g) Magnified view of the boxed region in (d), which compares the result of polarization modulation and pSIM. (h) The intensity profile of the line indicated in (e), whose Fourier transform (i) shows 184-nm periodicity, consistent with previous results. (j, k) The ¡°actin ring¡± structure is critical in Membrane-associated Periodic Skeleton (MPS) in neurons. The previous model assumes the ¡°end-to-end¡± organization of the Adducin capped actin filaments. However, pSIM reveals that the orientation of the short actin filaments is parallel to the axon shaft, supporting the ¡°side-by-side¡± organization of the ¡°actin ring¡± structure. Scale bars: (a) 2 $\mu$m (d-g) 1 $\mu$m.}
\end{figure}

\newpage
\begin{figure}[H]
\includegraphics[width=5in]{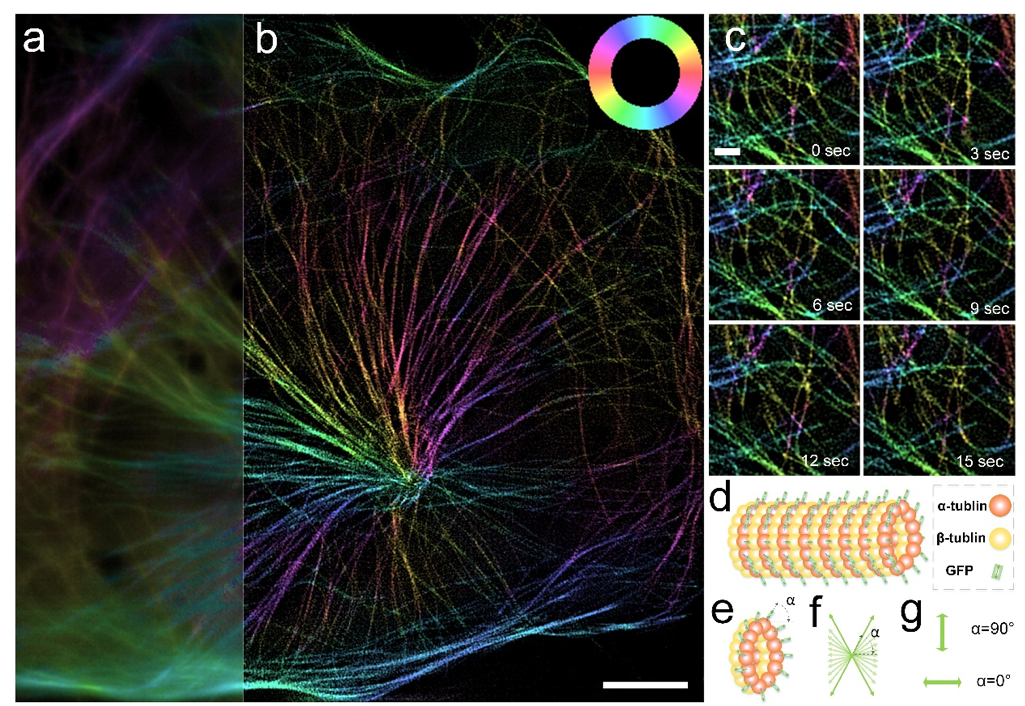}
\caption{\textbf{Live cell imaging microtubule in U2-OS cells.} (a, b) 3D-pSIM images the microtubule in a live U2-OS cell (80$\times$80$\times$0.75 $\mu$m$^3$) expressing tubulin-GFP at 0.67 reconstructed f.p.s. The M.I.P. images of polarization modulation (a) and pSIM (b) are compared. 3D-pSIM observes the in-plane dipole orientation in together with doubled lateral and axial resolution. (c) A time-lapse 2D section of the 3D-pSIM result. (d) The schematic of the $\alpha$-tubulin-GFP structure. The orientation of the ensemble dipole is perpendicular to the microtubule filament, (e-g) The ensemble dipole consists of 2D in-plane projection of all the 3D GFP dipoles. The relative orientation between the GFP dipole and the ¦Á-tubulin monomer determines whether the ensemble dipole is perpendicular or parallel to the filament. If the included angle ¦Á between the GFP dipole and the microtubule filament is close to 90$^\circ$ (or 0$^\circ$), the averaged dipole orientation is perpendicular (or parallel) to the filament. Scale bar: (a) 5 $\mu$m, (b) 1 $\mu$m. }
\end{figure}


\bibliographystyle{unsrt}


\end{document}